\documentstyle[11pt,a4,cite,epsf]{article}

\newcommand{\be}{\begin{equation}}
\newcommand{\ee}{\end{equation}}
\newcommand{\bea}{\begin{eqnarray}}
\newcommand{\eea}{\end{eqnarray}}
\newcommand{\noi}{\noindent}
\newcommand{\nn}{\nonumber}

\begin{document}

\begin{titlepage}
\begin{flushright} CPT-94/PE.3131\\ CERN-TH.7545/94
\end{flushright}
\vspace{2cm}
\begin{center} {\Large \bf On the large-$N_c$ behaviour of
the $L_{7}$ coupling in $\chi$PT.}\\[1.5cm]
{\large {\bf Santiago Peris}$^{a,b}$\footnote{On leave from Grup de Fisica
Teorica and IFAE, Universitat Autonoma de Barcelona, Barcelona, Spain.
E-mail: PERIS @ SURYA11.CERN.CH}\footnote{Work partially supported by
research project CICYT-AEN93-0474.} and
{\bf Eduardo de Rafael}$^a$}\\[1cm]
${}^a$  Centre  de Physique Th\'eorique\\
       CNRS-Luminy, Case 907\\
    F-13288 Marseille Cedex 9, France\\[0.5cm]
and\\[0.5cm]
$^b$ Theoretical Physics Division, CERN,\\
     CH-1211 Geneva 23, Switzerland.\\
\end{center}
\vspace*{1.5cm}
\begin{abstract}

It is shown that the usual large-$N_c$ counting of the coupling constant
$L_{7}$ of the  ${\cal
O}(p^4)$ low-energy chiral $SU(3)$ Lagrangian \cite{GL85} is in conflict
with general properties of QCD in the large-$N_c$ limit. The solution of
this conflict within the framework of a chiral $U(3)$ Lagrangian is
explained.

\end{abstract}
\vfill \begin{flushleft} December 1994\\
\end{flushleft} \end{titlepage}

{\bf 1.}
Chiral perturbation theory ($\chi PT$) is the effective field theory
of quantum chromodynamics (QCD) at low energies.
It describes the strong interactions of the low-lying pseudoscalar
particles in terms of the octet of Nambu-Goldstone fields
($\vec{\lambda}$ are the eight $3\times 3$ Gell-Mann matrices):

\be \label{eq:phi}
\Phi(x)={\overrightarrow{\lambda}\over
\sqrt{2}}\cdot\overrightarrow{\varphi}(x)=
\normalbaselineskip=18pt\pmatrix{\pi^{\circ}/\sqrt{2}+
\eta/\sqrt{6}&\hfill\pi^+\hfill &\hfill K^+\hfill\cr
\hfill\pi^-\hfill &-\pi^{\circ}/\sqrt{2}+\eta/\sqrt{6}&
\hfill K^{\circ}\hfill \cr
\hfill K^- \hfill &\hfill\overline K^{\circ} \hfill
&-2\eta/\sqrt{6}},
\ee
\noi
as explicit degrees of freedom, rather than in terms of the quark
and gluon fields of the usual QCD Lagrangian. In the conventional
formulation, the Nambu-Goldstone fields are collected in
a unitary $3\times 3$ matrix ${\cal U}(x)$ with $\det{\cal U}=1$,
which under $SU(3)$-chiral transformations $(V_L,V_R)$ is chosen to
transform linearly

\be
{\cal U}\rightarrow V_R{\cal U}V_L^{\dag}.
\ee

\noindent
The possible terms in the effective Lagrangian ${\cal L}_{eff}$,
with the lowest chiral dimension i.e., ${\cal O}(p^2)$ are
\cite{We79}:

\be \label{eq:lag2}
{\cal L}_{eff}^{(2)}={1\over 4}f^2_{\pi}\left\{tr \partial _{\mu}{\cal
U}\partial ^{\mu}{\cal U}^{\dag} +tr(\chi{\cal U}^{\dag} +{\cal
U}\chi^{\dag})\right\}.
\ee

\noi
The term with the matrix $\chi$ is the lowest order term induced by
the explicit breaking of the chiral symmetry in the underlying QCD
Lagrangian, due to the quark masses. For our purposes, it will be
sufficient to consider the approximate case where $m_u=m_d=0$. Then

\be
\chi\simeq {\rm diag}[0,0,2M_K^2].
\ee

\noi
An explicit representation of ${\cal U}$ is

\be
{\cal U}(x)=\exp\left(-i{1\over
f_{\pi}}\overrightarrow{\lambda}\cdot
\overrightarrow{\varphi}(x)\right),
\ee

\noi
and our normalization is such that $f_{\pi}=92.5\ MeV$.

The identification of all the independent local terms of ${\cal
O}(p^4)$, invariant under parity, charge conjugation, and local
chiral-$SU(3)$ transformations, as well as the phenomenological
determination of the ten physical coupling constants which appear,
has been made by Gasser and Leutwyler \cite{GL84}, \cite{GL85}. We
reproduce the terms which will be relevant for our discussion:

\be \label{eq:lag4}
{\cal L}_{eff}^{(4)}\doteq L_{5}tr[\partial_{\mu}
{\cal U}^{\dagger}
\partial^{\mu}{\cal U}
(\chi^{\dagger}{\cal U}+{\cal U}^{\dagger}\chi)]
+L_{7}[tr(\chi^{\dagger}{\cal U}-{\cal U}^{\dagger}\chi)]^2+
L_{8}tr(\chi^{\dagger}{\cal U}\chi^{\dagger}{\cal U}+
\chi{\cal U}^{\dagger}\chi{\cal U}^{\dagger}).
\ee

\noi
Notice that in the limit $m_u=m_d=0$ there are no ambiguities of the
type observed by Kaplan and Manohar \cite{KM86}.

It is well known that in the large-$N_c$ limit of QCD \cite{'tH74},
the constant
$f_{\pi}$ is ${\cal O}(\sqrt{N_c})$. In
ref.\cite{GL85} it was shown that
$L_{5}$ and $L_{8}$ are of ${\cal O}(N_c)$, while the
conclusion for $L_{7}$ was that it should be considered as of ${\cal
O}(N_c^2)$. The purpose of this note is to discuss some implications of
the large-$N_c$
counting of the $L_{7}$ constant. Our arguments implicitly assume
that general properties derived from QCD in the large-$N_c$ limit
hold order by order in $\chi PT$.
\newpage
{\bf 2.}
Expanding the Lagrangians ${\cal L}_{eff}^{(2)}$ and ${\cal
L}_{eff}^{(4)}$ in powers of the Nambu-Goldstone fields
$\overrightarrow{\varphi}(x)$ gives a string of interaction terms.
However, to leading order in the limit $N_c\rightarrow \infty$, and
due to the fact that the
$\overrightarrow{\varphi}(x)$-fields are always normalized to
$f_{\pi}$ -which itself is ${\cal O}(\sqrt{N_c})$-, only kinetic-like
terms and mass-like terms survive from the expansion
in ${\cal L}_{eff}^{(2)}$. With $L_{5}$ and $L_{8}$ considered as of
${\cal O}(N_c)$, the same happens with the terms induced by these
couplings. It is easy to check that the omitted couplings in  ${\cal
L}_{eff}^{(4)}$, all lead to terms which vanish in the
strict  $N_c\rightarrow \infty$ limit. This property is in fact
in agreement with general arguments which assert that QCD in
the  $N_c\rightarrow \infty$ limit is a theory of
non-interacting mesons \cite{W79b}.

The terms generated by the
expansion of the trace modulated by the $L_{7}$ coupling in
eq.(\ref{eq:lag4}) require special attention. When restricted to terms
relevant to the purpose of the discussion here, one finds

\bea
L_{7}[tr(\chi^{\dagger}{\cal U}-{\cal U}^{\dagger}\chi)]^2 &
= & -L_{7}\frac{64}{3}\frac{M_K^4}{f_{\pi}^2}\eta(x)\eta(x)
\nn \\
& & -L_{7}\frac{64}{3\sqrt{3}}\frac{M_K^4}{f_{\pi}^4}
\eta(x)\pi^0(x)K^{+}(x)K^{-}(x)+ \cdots,
\eea

\noi
where the dots denote other 4-meson interactions. With
$L_{7}$ considered as of ${\cal O}(N_c^2)$ in the large-$N_c$
limit we are confronted with (at least) the following serious
problems:
\begin{description}

\item[i)]
The quadratic term, a mass like term, diverges in the
$N_c\rightarrow \infty$ limit contrary to the expected constant
behaviour \cite{W79b}. Furthemore, with
$L_{7}<0$
 -as suggested from phenomenology \cite{GL85}-  it is
tachyonic!

\item[ii)]
The quartic $\eta\pi^{0}K^{+}K^{-}$ term which remains in the limit
$N_c\rightarrow \infty$
plays the r\^ole of a
$\lambda \varphi^4$-like interaction with a non-vanishing negative
($L_{7}<0$) coupling, again in contradiction with the expected
non-interacting behaviour \cite{W79b}. This would also imply that the
effective potential is unbounded from below and the theory, in the
limit
$N_c\rightarrow
\infty$, becomes unstable!

\end{description}

In view of these conflicts, it seems mandatory to reconsider the reasons
which lead to considering $L_7$ as of ${\cal O}(N_c^2)$ in the
large-$N_c$ limit.

\vspace{7 mm}
{\bf 3.}
The reason why it is usually assumed that $L_7\sim {\cal O}(N_c^2)$
is because of the contribution of the $SU(3)$-singlet, the $\eta_0$.
Indeed, when discussing the large-$N_c$ limit, it is convenient to
work with the
$U_L(3)\times U_R(3)$ effective Lagrangian which includes nine
Nambu-Goldstone fields. To leading order in the chiral expansion and
in the
$1/N_c$-expansion the Lagrangian is (see refs.
\cite{W79} to \cite{SSW93}):

\be\label{eq:eight}
{\cal L}(\tilde{\cal U})=
{1\over 4}f^2_{\pi}\left\{tr \partial _{\mu}\tilde{\cal U}\partial
^{\mu}\tilde{\cal U}^{\dagger} +tr(\chi\tilde{\cal U}^{\dagger}
+\tilde{\cal U}\chi^{\dagger}) +
\frac{a}{4N_c}(tr\log\frac{\tilde{\cal U}}
{\tilde{\cal U}^{\dagger}})^2 \right\},
\ee
\noi
where

\be
\tilde{\cal U}=\exp (-i\sqrt{2/3}\frac{\eta_{0}(x)}{f_{\pi}})\,
{\cal U}.
\ee

\noi
The constant $a$ has dimensions of mass squared, and with the
$1/N_c$ factor pulled out, it is of ${\cal O}(1)$ in the
large-$N_c$ limit. At the same level of approximations, the
expansion in powers of the
$\eta_{0}$ field results in the expression:

\bea \label{eq:lageta}
{\cal L}(\tilde{\cal U}) & = &
\frac{1}{2}\partial_{\mu}\eta_0\partial^{\mu}\eta_0-
\frac{1}{2}(\frac{3a}{N_c}+\frac{2}{3}M_K^2)\eta_0\eta_0 \nn \\
& & -i\sqrt{2/3}\frac{f_{\pi}}{4}\eta_{0}tr(\chi^{\dagger}{\cal
U}-{\cal U}^{\dagger}\chi)+{\cal L}_{eff}^{(2)},
\eea

\noi
where ${\cal L}_{eff}^{(2)}$ is the same as in eq.(\ref{eq:lag2}),
and therefore has no $\eta_{0}$ field couplings.

Integrating out the $\eta_0$ field results in general in a non-local
interaction of the form

\be \label{eq:eleven}
f_{\pi}^2\ \int d^4xd^4y
[tr(\chi^{\dagger}{\cal U}(x)-{\cal U}(x)^{\dagger}\chi)]\ \ D(x-y)\ \
[tr(\chi^{\dagger}{\cal U}(y)-{\cal U}(y)^{\dagger}\chi)]\ ,
\ee
with
$$
\int d^4x\ e^{ipx}\ D(x)=\left[ p^2 -
\left(\frac{3a}{N_c}+\frac{2}{3}M_K^2\right)\right]^{-1}\ \ .
$$
As long as one keeps $N_c$ {\sl finite}, and to the extent that
$3a/N_c>>M_K^2$, one can envisage an expansion in powers of momentum that
yields a tower of {\sl local} operators. To lowest order in this expansion
one finds
\be \label{eq:twelve}
{\cal L}(\tilde{\cal U})\Rightarrow {\cal L}_{eff}^{(2)}-
\frac{f_{\pi}^2}{48(\frac{3a}{N_c}+\frac{2}{3}M_K^2)}
[tr(\chi^{\dagger}{\cal U}-{\cal U}^{\dagger}\chi)]^2.
\ee
There appears then an $L_7$ term, with an estimate for the induced coupling
constant :

\be \label{eq:thirteen}
L_7^{\eta'}=- \frac{f_{\pi}^2}{48(\frac{3a}{N_c}+\frac{2}{3}M_K^2)}.
\ee

\noi
In terms of physical masses: $\frac{3a}{N_c}+\frac{2}{3}M_K^2\simeq
M_{\eta}^2+M_{\eta'}^2-\frac{4}{3}M_K^2\simeq M_{\eta'}^2$, and
$L_7^{\eta'}\simeq -2\times 10^{-4}$.

Let us now discuss the large-$N_c$ limit. Taking the limit $N_c\rightarrow
\infty$ on the expression (\ref{eq:twelve}) invalidates the condition under
which eq. (\ref{eq:twelve}) was obtained. If, in spite of this fact, one still
takes this limit one finds that the answer crucially depends on whether one
takes the chiral limit first and $N_c\rightarrow \infty$ afterwards or the
other way around. The usual result $L_7\sim {\cal O}(N_c^2)$ comes from
first neglecting $M_K^2$ in eq. (\ref{eq:thirteen}) and then taking
$N_c \rightarrow \infty$. This faces the problems with the large-$N_c$
counting of QCD that we mentioned at the beginning. If, on the contrary, one
takes the limit $N_c\rightarrow \infty$ keeping $M_K^2$ finite one finds
that the chiral counting is upset and one can no longer consider $L_7$
(which now would be of ${\cal O}(N_c)$ instead) as a coefficient of the
${\cal O}(p^4)$ chiral $SU(3)$ Lagrangian.
Of course both  situations stem from the fact that in the limit
$N_c\rightarrow \infty$ the interaction (\ref{eq:eleven}) cannot be considered
local and therefore, strictly speaking, it cannot be encoded into an $L_7$
term. The limit $N_c\rightarrow \infty$ has to be described by enlarging the
chiral $SU(3)$ group to chiral $U(3)$ (i.e. the Lagrangian of eq.
(\ref{eq:eight}) plus higher order terms). There will also be an ${\cal
O}(p^4)$ $L_7$-{\sl like} term in this chiral $U(3)$ effective Lagrangian,
but it will be at most of ${\cal O}(N_c)$ at large $N_c$ since the $\eta_0$
is an explicit field in the Lagrangian. Then no inconsistencies arise.

There is, however, a sense in which taking the limit $N_c\rightarrow \infty$
in eq. (\ref{eq:thirteen}) is still meaningful. This is when going from
$U_L(3)\times U_R(3)$ to the limit $U_L(2)\times U_R(2)$. In this case the kaon
is
no longer a Nambu-Goldstone particle and can be treated as a heavy particle
in an effective theory (of two light flavours) with momenta $p^2<<M_K^2$.
Then $M_K^2$ in eq. (\ref{eq:thirteen}) is kept finite and the mass term for
the $\eta$-field in the Lagrangian (\ref{eq:twelve}) has two sources, with the
result
\be
-\frac{1}{2}\ \frac{a}{N_c}\ \frac{4M_K^2}{\frac{3a}{N_c}+\frac{2}{3}M_K^2}\
\eta(x)\ \eta(x)\ \ .
\ee
In the limit $N_c\rightarrow \infty$ the Lagrangian (\ref{eq:twelve}) reveals
the existence of four Nambu-Goldstone particles: the three pions and the
$\eta$ singlet. With $m_u\not= m_d\not= 0$, the same Lagrangian describes
the effective theory of four (pseudo) Nambu-Goldstone bosons with an
explicit $L_7$-type interaction. The coupling constant of this interaction
term, which is the $U_L(2)\times U_R(2)$ equivalent of the $l_7$ in refs.
\cite {GL84}\cite{GL85}, appears then as of ${\cal O}(N_c)$ in the large-$N_c$
limit.

{}From the analyses above, we are led to the conclusion that, if the
low-energy effective field theory of QCD with three light flavours is to
remain compatible with the large-$N_c$ limit of QCD, a safe way to formulate
the effective chiral Lagrangian is within the framework of $U_L(3)\times
U_R(3)$
instead of $SU_L(3)\times SU_R(3)$. In that respect, a systematic study of the
phenomenological implications of low-energy hadron physics within that
framework, in particular in the sector of $\eta(\eta')$-decays and, perhaps,
non-leptonic $K$-decays, seems worthwhile.

\vskip 0.5in

{\bf Acknowledgements}: We are  grateful to J. Bijnens, G. Ecker, G.
Esposito-Far\`ese, J. Gasser, H. Leutwyler, A. Pich, J. Prades and G.
Veneziano for their comments and criticisms.

\newpage



\end{document}